\documentclass[prl,aps,amssymb,nofootinbib,twocolumn,showpacs]{revtex4}
\usepackage{amsmath}
\usepackage{amssymb}
\usepackage{amsthm}
\usepackage{amsfonts}
\usepackage{algorithmic}
\usepackage{enumerate}
\usepackage{latexsym}
\usepackage[dvips]{graphicx}

\newcommand{\beq}{\begin{equation}}
\newcommand{\eneq}{\end{equation}}
\newcommand{\beqs}{\begin{equation*}}
\newcommand{\eneqs}{\end{equation*}}

\input{epsf}

\begin{document}

\tolerance 10000

\title{The $SU(2)$ Gauge Theory of the Substantially Doped Mott-Insulator:
\\ the phase separated d+s superconductor.}

\author { B.W.A. Leurs and J. Zaanen}

\affiliation{ Instituut Lorentz for Theoretical Physics, Leiden University,
Leiden, The Netherlands}

\begin{abstract}
\begin{center}

\parbox{14cm}{We reformulate the mean-field theory associated with the
$SU(2)$ gauge theory of spin-charge separation in doped Mott-insulators of Wen and Lee so that
it can deal with the regime of substantial carrier density. We find that it describes
remarkably accurately phase separation tendencies. Moreover, for  elegant reasons the theory
insists that the superconducting state  should have a $d+s$ orderparameter with a s-wave
component growing in the overdoped regime. This appears to be consistent with both Raman- and tunneling measurements
in cuprate superconductors.}
\end{center}
\end{abstract}

\date{\today}

\maketitle

The d-wave nature of the superconducting order parameter in the cuprates is often taken as the
leading evidence for this superconductivity to be caused by an non-phonon mechanism rooted in
strong electron-electron interaction physics. However, a number of evidences
appeared in independent experiments indicating that the order parameter breaks the symmetry
further. The d-wave order parameter respects the rotational symmetry of the tetragonal
(square) lattice, but by the admixing of a s-wave component this is spontaneously broken to an
orthorombic symmetry.  In orthorombic crystals like the YBCO superconductors such
a s-wave admixture is automatic (and well established\cite{hilgenkamp}) but both in Raman-
\cite{tajima}and c-axis tunneling\cite{klemm} experiments evidences have been found for a
substantial (up to 20\%)  s-wave admixture to occur spontaneously in the BISCO superconductors
characterized by a square $Cu-O$ lattice.

 It seems that in none of the various candidate theories
for the mechanism of superconductivity one finds a truly good reason for $d+s$ symmetry. The 'slave'
$SU(2)$ gauge theory due to Wen, Lee and coworkers\cite{wenlee} is among the most sophisticated
unconventional high Tc theories. Motivated by the question of how to deal with charge inhomogeneity
in this theoretical setting we found out how to compute in the regime of substantial doping, staying strictly within the
original formulation of the theory. On the one hand, we find that the $SU(2)$ theory deals remarkably well
with phase separation, but we also discovered that for elegant reasons rooted in the projective symmetry
structure\cite{wenpsg} behind the theory, the superconductivity {\em has to be} $d+s$ in the overdoped regime.
Our main
findings are summarized in the figures. As function of increasing $J/t$ we find that  a phase separation regime
opens up at low doping (Fig. 1) saturating in a complete phase separation at $J/t \simeq 4$,
consistent with the outcomes
of exact numerical diagonalization studies\cite{hellberg,kivelsonPS}.
Also, the electronic incompressibility calculated for a realistic
$J/t = 0.3$ is remarkably consistent with experiment (Fig. 2). This suggests that by adding physics beyond
the $t-J$ model (e.g., long range Coulomb interactions) the $SU(2)$ theory can deal with the stripes
\cite{zaanen,white,kivelsonstripe,seamus}.
Above a doping $x=0.15-0.20$  a homogeneous superconducting phase with $d+s$ symmetry is predicted,
characterized by an $s$-wave component growing proportional to the doping and a $d$-wave component that is
decreasing, to vanish completely at a doping $x \simeq 0.3$ (inset Fig.1).

The above was overlooked in the earlier work\cite{wenlee,fatfollowuppapers} for mere technical reasons:
to simplify the calculations the focus was on the low
hole density limit where the charge sector can be described as a dilute bose gas.
However, at higher dopings one has to do justice to the hard-core nature of the bosonic
charge sector and this is actually quite easy to implement on the mean-field level.
Before we go into the details of the calculations let us first present a simple argument making
transparent why the $SU(2)$ projective symmetry  structure (i.e., the mean-field states span
up the gauge volume\cite{wenpsg}) implies that the superconducting order parameter has to be $d+s$.  The
idea to invoke an enlarged ($SU(2)$) gauge symmetry is rooted in the observation
 by Affleck and Marston\cite{affleck}
that the spin system at half-filling in a fermionic  'spinon' representation is
characterized by both the usual 'stay at home' $U(1)$ gauge symmetry, and a local charge
conjugation symmetry meaning that one can describe the spin system equally well in terms of
spinon particles and anti-particles. In this
representation the spin system is invariant under $SU(2)$ gauge transformations  $\psi_{i}
\rightarrow g_{i}\psi_{i}$ where $\psi_{i} = (f_{\uparrow i}, f^{\dag}_{\downarrow i})$  (with
$f^{\dagger}$ creating a spinon) and $g_{i} \in SU(2)$. Away from half-filling the charge
conjugation is broken and $SU(2) \rightarrow U(1)$. However, the $SU(2)$ symmetry is only weakly
broken and the gauge transformations away from the 'equator of the isospin
sphere' turn into soft physical modes\cite{fatfollowuppapers}. To encode this mathematically,
Wen and Lee introduced a doublet of bosonic holons $h_i = (b_{1i} , b_{2i})$ transforming covariantly
as $h_{i} \rightarrow g_{i}h_{i}$. To appreciate the meaning of the boson $SU(2)$ flavors it is
informative to inspect how the physical (gauge invariant) Hilbert
space looks like\cite{wenlee}.  The singly occupied states correspond with
one spinon per site $f^{\dag}_{\sigma}\left|0\right>$, while
the physical hole corresponds with the interesting combination
$\frac{1}{\sqrt{2}}(b^{\dag}_1 + b^{\dag}_2
f^{\dag}_{\downarrow}f^{\dag}_{\uparrow})\left|0\right>$. This reveals
the meaning of the boson doublet. The hole is a spin-singlet but
the $SU(2)$ parametrization remembers the charge conjugation
symmetry of half-filling, insisting that singlets formed from missing
spinons or of local {\em s-wave} pairs of spinons are gauge
equivalent. The physical hole is a coherent superposition of these
two possibilities and the holon flavor takes care that the state as
a whole is a gauge singlet.

On the mean-field level the spinon sector takes care of the symmetry of the superconducting
order parameter. Let us consider the completely empty state: the mean-field theory is of course
unphysical because the gauge fluctuations will have destroyed the deconfining spin-charge separated
state already at much lower hole densities. But it is an informative exercise. The exact wave function
 of the empty state is $ | vac
\rangle_{phys} = \Pi_i \left( b^{\dagger}_{i1 } +  b^{\dagger}_{i2} f^{\dagger}_{i \downarrow}
f^{\dagger}_{i \uparrow} \right) | vac \rangle_{unphys} $. Spin-charge
separation implies that the system looses its knowledge about the three particle correlation
$\sim b^{\dagger}_{i2} f^{\dagger}_{i\downarrow} f^{\dagger}_{i \uparrow} $. In addition, one
has to satisfy the local constraint
$\left(\psi^{\dag}_{i}\tau^l \psi_{i} +
h^{\dag}_{i}\tau^l h_{i}\right)\left|\mbox{phys}\right> = 0$,$l=1,2,3$,  implying
that the bosons have a hard core: at most one boson of either flavor can be
present at a single site.  The best choice for a holon-spinon product wave function becomes obvious:
$
 | vac \rangle_{MF} = \Pi_i \left[ \frac{1}{2} \left( b^{\dagger}_{i1 } + b^{\dagger}_{i2}
\right) \left( 1 + f^{\dagger}_{i \downarrow} f^{\dagger}_{i \uparrow} \right) \right] | vac
\rangle_{unphys}.
$
 In this mean field state there is exactly one boson present and it
follows that this state decribes a holon Mott-insulator together with a local pair s-wave
spinon superconductor. Of course, this state is nonsensical ('nothingness' is not a
superconductor) but it does remarkably well as a density functional theory\cite{kohn} ground state.
The energy of the 'exact' empty state is of course vanishing and the same is true for the above
mean field state: the hard-core of the bosons kills the hopping energy, and
since the spin-operators have the form $S^{l}
=f^{\dag}_{\alpha}\sigma^{l}_{\alpha\beta}f_{\beta}$, these vanish as well. Lastly, it is readily checked
that the three projection operators $\psi^{\dag}_{i}\tau^{l} \psi_{i} +
h^{\dag}_{i}\tau^{l} h_{i}$ also act like zero. In what follows we will merely use the slave mean-field
theory as a density functional theory and the 'magic' revealed in the above gives us reasons to believe
that it might work better than naively expected.

Let us now turn to the full mean-field theory. The starting point is
the usual $t-J$ model and after the Hubbard-Stratonovich
decoupling in the spinon sector in terms of  the auxiliary fields
$\chi_{ij}=\left<
f^{\dag}_{\uparrow i}f_{\uparrow j} +
 f^{\dag}_{\downarrow i}f_{\downarrow j}\right>$ and
$\Delta_{ij} = \left< f_{\uparrow i}f_{\downarrow j} - f_{\downarrow i}f_{\uparrow j}\right> $
one obtains the standard result,
\begin{eqnarray} \label{mftheory}
H_{mf} =  \sum_i -\mu \  h^{\dag}_{i}h_{i} - a_{0i}^{l}
\left(\frac{1}{2}\psi^{\dag}_{\alpha
i}\tau^{l}\psi_{\alpha i} + h^{\dag}_{i}\tau^{l} h_{i} \right) \\
 + \sum_{< ij >} \frac{3J}{8}\left(\mbox{Tr}U_{ij}^{\dagger}U_{ij}
              +\psi^{\dag}_{i}U_{ij}\psi_{j} \right) +
 t h^{\dag}_{i}U_{ij}h_{j}  + h.c.   \nonumber
\end{eqnarray}
The mean fields are grouped as $
 U_{ij} =  \left (
      \begin{matrix}
   -\chi_{ij}^*   & \Delta_{ij} \\
    \Delta_{ij}^* & \chi_{ij}
      \end{matrix}
      \right)
$, and $iU_{ij}\in SU(2)$, up to a scalar factor.
The mean-fields are invariant under the projective transformation
 $ U_{ij} \rightarrow \tilde{U}_{ij}= g_i U_{ij}
g^{\dag}_j$,$ \tilde{a}^{l}_{0i}\tau^{l} = g^{\dag}_{i}a^{l}_{0i}\tau^{l}g_i .$ This implies
that we may choose some reference state $U_{ij}^{0}$, and label all other mean-field states by
$g_i$. We choose as reference mean-field state the d-wave superconducting state (dSC),
$U^{0}_{i,i+\hat{x}}=-\chi\tau^3 + \Delta\tau^1 , U^{0}_{i,i+\hat{y}} =-\chi\tau^3 -
\Delta\tau^1$.
The mean-field solutions in the projective gauge volume\cite{wenpsg}
are characterized by the isospin $SU(2)$ sphere; we take
 $ g_{i} = \exp\left(-i\frac{\theta^{F}_i}{2} \tau^2\right) $
such that the fermionic latitude angle $\theta^{F} =\pm \frac{1}{2}\pi$ decribes the staggered flux phases
and $\theta^{F} = 0$ the superconductor, while the longitude ($\varphi$) represents the
$U(1)$ gauge\cite{wenlee}.

In Eq. (\ref{mftheory}) the holons are still fully dynamical. However,
we learned from the empty limit that their hard-core nature is
crucial and it is quite simple to construct
condensate wavefunctions for hard-core bosons.These are
just like XY spins and the mean-field states have the generalized
coherent state form, taking care of the $SU(2)$ doublet:
\begin{equation}
| \Psi_0 \rangle_{holons} = \Pi_i \left( \alpha_i + \beta_{i} (u_i   b^{\dagger}_{i1 } + v_i
b^{\dagger}_{i2 } ) \right) | vac \rangle_{unphys}
\label{holonMFstate}
\end{equation}
where
 $u_i = \cos(\frac{\theta_i}{2}), v_i = \sin(\frac{\theta_i}{2})$, and $|\alpha|^2 = 1 -
 |\beta|^2$.
 Further,
$\beta_i \rightarrow \beta_{0i} e^{i\varphi_i}$ such that $\beta_{0i}$ is real in terms isospin angles
 $\theta, \phi$. For clarity's sake, we distinguish the bosonic angle $\theta$ from the
 fermionic isospin angle $\theta^{F}$.
  This parametrisation is convenient, since $\theta_i = \frac{\pi}{2}$ makes the expectation
values for $b_1$ with vacancies indistinguishable from $b_2$ with a spinon pair, reproducing
the particle-hole symmetric empty state $\left| vac \right>_{phys}$.
On the mean-field level, the local constraint equation  is replaced by its average:
 $ \left<
\psi^{\dag}_{i}\tau^l \psi_{i} + h^{\dag}_{i}\tau^{l}h_{i} \right> = 0, $ and for
the condensate Ansatz Eq. (\ref{holonMFstate}) this becomes
\begin{eqnarray}
\left< f^{\dag}_{\uparrow i} f^{\dag}_{\downarrow i} + f_{\downarrow i} f_{\uparrow i} \right>
&=& |\beta_{0i}|^{2} \sin(\theta_i)\cos(\varphi_i)
\label{constr1} \\
-i \left< f^{\dag}_{\uparrow i} f^{\dag}_{\downarrow i} - f_{\downarrow i} f_{\uparrow i} \right>
&=& -i |\beta_{0i}|^{2} \sin(\theta_i)\sin(\varphi_i)
\label{constr2}
\\
\left< f^{\dag}_{\alpha i}
f_{\alpha i} -1 \right> &=& |\beta_{0i}|^{2} \cos(\theta_i).
\label{constr3}
\end{eqnarray}
Eq. (\ref{constr2}) is satisfied by  the $U(1)$ gauge $\varphi_i=0$, such that $a^2_{0i}=0$, and
 Eq. (\ref{constr3})
is standard. The fermionic expectation values are calculated with respect to the dSC gauge.
However, in the original treatment by Wen and Lee\cite{wenlee}, as well as in the follow up
work\cite{fatfollowuppapers} the first constraint equation Eq. (\ref{constr1}) appeared to be
automatically satisfied as a consequence of the treatment of the holons in terms of the dilute
bose-gas limit. This is not at all true at substantial dopings, as can be immediately seen
from Eq. (\ref{constr1}): for any finite doping ($|\beta_{0i}|\not=0$), in the presence of a
d-wave superconducting order
 parameter ($\theta=\frac{\pi}{2}$), one \textit{needs} to have a finite s-wave component that is
 linearly proportional to the doping: $<f^{\dag}_{\uparrow i} f^{\dag}_{\downarrow i} > = \frac{x}{2}$!
 The presence of d-wave superconductivity is of course not necessary at all for the s-wave superconductor,
 as is immediately obvious from the empty limit.

Let us now explicitly derive the free energy functional from the Hamiltonian (\ref{mftheory}).
We have checked that at least for the $t-J$ model homogeneous states are preferred and here we
specialize to $\theta_i = \theta, |\beta_{0i}| = \beta$ and $a^{l}_{0i}=a^{l}_{0}$, while
$\chi_{ij}$ and $\Delta_{ij}$ are chosen to describe a homogeneous flux phase and/or d-wave
superconductors. The  ground state energy per site $e_{MF}$ becomes
\begin{eqnarray}\label{mfielden}
e_{MF} & = & -\frac{1}{N} \sum_k E_k +  \frac{3}{4 N}J( |\chi|^{2} + |\Delta|^{2})
\\
& & - \; 2 t \chi |\alpha|^2 |\beta|^2 - (\mu + a_0^1 \sin \theta + a^3_0 \cos\theta
) |\beta|^2 \nonumber
\end{eqnarray}
Here $ E(k) = \sqrt{ (\chi_k - a^3_0)^2 + (\Delta_k - a^1_0)^2 } $ is the spinon dispersion
relation, with $\chi_k = \frac{3J}{4}\chi (\cos k_x + \cos k_y)$ and $\Delta_k =
\frac{3J}{4}\Delta(\cos k_x - \cos k_y)$. Notice that we consider the mean-field energy density functional
in the grand canonical ensemble, by fixing
$\mu$ instead of the density $\rho$, anticipating phase separation.
Using the notation $\rho = |\beta|^2 $ , the four saddle point equations are
\begin{eqnarray} \label{mfeq}
2 \chi  &=& \frac{1}{\chi} \sum_k \frac{\chi_k^{2}}{\ E_k}+2\left\{ \frac{4 t1}{3 J}
\frac{\partial}{\partial \chi}
 \rho(\chi) (1 -\rho(\chi)\ ) \right\},
\\
2 \Delta  &=& \frac{1}{\Delta}\sum_k \frac{(\Delta_k -a^{1}_0)\ \Delta_k}{ \ E_k}, \mbox{\ \ \ }
\rho(\chi) = \sum_k \frac{(\Delta_k -a^{1}_0)}{E_k} \nonumber
\\
0 &=& \rho(\chi)\left\{ \rho(\chi) - \frac{1}{2}\left( 1 + \frac{\mu + a^{1}_{0}\sin\theta +
a^{3}_{0}\cos\theta}{2t\chi} \right)  \right\} . \nonumber
\end{eqnarray}
where the third equation is the homogeneous form of Eq. (\ref{constr1}). We find  solutions by
minimizing the energy (\ref{mfielden}) numerically with the simulated annealing method
\cite{simann}. We also calculated the saddle point solutions for a staggered $a^{1}_{0i}$, to
incorporate the possibility of the staggered flux state \cite{affleck}. Due to Eq.
(\ref{constr1}), a staggered $a^{1}_{0i}$ leads to a staggered $s$-wave component.

\begin{figure}[ht]
\centering \rotatebox{0}{
\resizebox{5.8cm}{!}{%
\includegraphics*{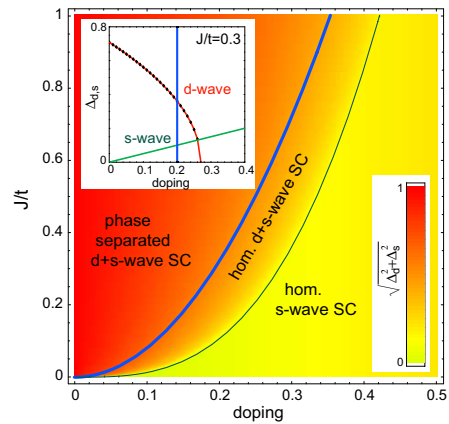}}}
\caption{ The zero temperature phase diagram for the  $SU(2)$-mean field theory for the $t-J$ model,
as function of doping $x$ versus the ratio of exchange ($J$)  and hopping ($t$) energies.
There is a regime to the
left of the blue line where the system macroscopically phase separates in the Mott-insulator and a
homogeneous $d+s$ wave superconductor. This $d+s$ superconductor only exists in a rather small
doping regime adjacent to the phase separation line, and a transition follows to a pure s-wave
superconductor at higher doping (green line, see the inset for the doping dependence of the gaps
$\Delta_{d,s}$ for a 'physical' $J/t =0.3$).
The false colors code for the overall gap magnitude, and these together with the
$d$ and $s$ gaps in the inset are computed in the canonical ensemble in the phase separation regime,
to indicate how they would evolve when phase separation would not take place.}
\end{figure}

Some main features survive in this revised version of the mean-field theory: in the presence
of holes the superconducting state is always preferred over the flux phase\cite{kotliar,wenlee}
, and at least for the $t-J$ model it appears that uniform states have always
 a better energy than inhomogeneous states\cite{tobepublished}. A first novelty is that
 we now find that macroscopic phase separation occurs generically at low hole density:
 as function of chemical potential, the system stays initially at half-filling and pending the
 ratio of $J/t$
at some finite $\mu$ a level crossing takes place to a state with a finite doping level  (Fig.1 ). As
function of increasing $J/t$ the width of this phase separation regime is increasing (see Fig.
1) and we find that for $J/t \simeq 4$ the phase separation is complete,
consistent with exact diagonalization studies on the t-J model indicating a complete phase
separation for $J/t \geq 3.5$ \cite{hellberg,kivelsonPS}. This is by itself a rather serious test
of this mean-field theory in its capacity of a density functional theory:
to reproduce the critical $J/t$ for complete phase separation, both the energies of the Mott-insulator
 ($x=0$)
and of the states having a hole density $x \neq 0,1$ have to be quite accurate, given that the mean-field
energy of the empty state ($x=1$) is exact. How well does this compare to experiment? In Fig. 2 we
compare
the calculated doping dependence of the chemical potential for a 'physical' $J/t =0.3$  with the
 photoemission
results  by Fujimori and coworkers\cite{fujimori}. The doping independence of $\mu$ is indicative of phase
separation and the theory predicts a wider phase separation regime ($x_{crit} \simeq 0.2$) than experiment
($x_{crit} \simeq 1/8$): this discrepancy has likely more to do with the $t-J$ model itself than with the
 mean-field
theory. However, we find that the electronic incompressibility $1 /\kappa =
\frac{\partial^{2}E_{\mbox{\scriptsize{MF}}}}{\partial x^{2}} = \frac{\partial\mu}{\partial x
}$ of the homogeneous superconductor at higher dopings is remarkably well reproduced
 by the theory (Fig. 2).

\begin{figure}[ht]
\centering \rotatebox{0}{}
\resizebox{6.3cm}{!}{%
\includegraphics*{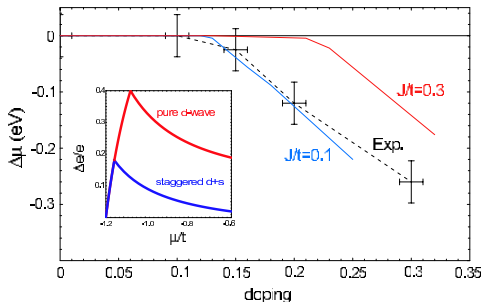}}
\caption{ The chemical potential shift as a function of doping computed for $J/t = 0.1, 0.3$.
Using $t = 0.44$ eV \cite{hybertsen} the absolute values of $\Delta \mu$ can be compared with
the experimental (photoemission) results by  Fujimori {\em et al.}\cite{fujimori}: taking the
'physical' $J/t = 0.3$ the theory overestimates the critical doping where phase separation
stops, but the theory is accurately reproducing the electronic incompressibility (the slope of
the lines) of the homogeneous superconductor. The inset shows the relative mean-field energy
cost of both the uniform pure $d$ wave superconductor (red) and the superconductor with
staggered $s$-wave component (blue), relative to the homogeneous $d+s$ superconducting ground
state as a function of chemical potential. The spikes in the relative energy differences are
due to the fact that for low chemical potential, the pure $d$-wave and the staggered
$d+s$-wave superconductors do not admit a non-zero holon density. }
\end{figure}

Let us now address the nature of the order parameter of the homogeneous superconductors
found at higher dopings (inset Fig. 1). We already established that the s-wave component
$<f^{\dag}_{\uparrow i} f^{\dag}_{\downarrow i} > = \frac{x}{2}$, as a direct consequence of
the constraint  Eq. (\ref{constr1}). To further emphasize this
matter, we compare in the inset of Fig. 2 the energy of a state where we have fixed the
Lagrange multiplier $a^{1}_{0}=0$ such that the s-wave component vanishes, with the best $d+s$ state,
finding that the former is indeed a false vacuum. The surprise is that according to the $SU(2)$ mean-field
theory a phase transition is present at a higher doping where the $d$-wave component
{\em vanishes altogether}.
As can be seen from Fig. 1, a d-wave component is only found in a relatively narrow doping region
 adjacent
to the phase separation line, regardless the $J/t$ ratio. Notice that we show in Fig. 1 the
development of the $d$ and $s$ wave components and the overall gap magnitude
as computed in the canonical ensemble: because we impose an
average density these are actually unphysical in the phase-separation regime.

Summarizing, by just improving the technical formulation of the $SU(2)$ gauge theory on the mean-field
level (holons as hard-core bosons) we have discovered that this theory exerts some remarkable, unexpected
powers. Viewed as a density functional theory\cite{kohn}, it is quantitatively accurate, even in a
regime where its
literal interpretation as a dynamical theory becomes nonsensical
(the confinement problem at large doping).
It reproduces the macroscopic phase separation tendency of the $t-J$ model accurately and it is well
understood that this is an unphysical pathology of the this model itself. It just signals the presence
 of stripe-like
instabilities\cite{zaanen,white,kivelsonstripe,seamus} and we leave it for future work to find out what
the
gauge theory has to tell about stripes.

The punchline is that the theory produces a strong- and counterintuitive prediction that
is well in reach of experimental confirmation: for deep and elegant reasons specific for the
workings of the underlying projective gauge principle\cite{wenpsg},
the superconducting order parameter in the overdoped regime
has to be $d+s$, with an s-wave component increasing {\em linearly} with doping and a decreasing d-wave
component, that is governed by the proximity of a transition to a pure s-wave state at a doping level
which is likely
beyond the level that can be experimentally realized.  This is surely not falsified by the present state
of
experimental affairs. We predict a substantial s-wave admixture at optimal doping of order
$10-20 \% $, which is consistent with $c$-axis tunneling experiments \cite{klemm}. We predict that the gap
ratio $r= \Delta_s/\Delta_d  $ grows with doping,  in accord with the Raman measurements \cite{tajima}.
There are quite a number of other, in principle straightforward ways to measure the evolution of the $d+s$
order parameter and we hope that the opportunity to (dis)prove the Wen-Lee gauge theory will be sufficient
motivation for such an experimental effort.

\noindent \textbf{ Acknowledgements:} We thank O. Vafek, M.S. Golden, S. Tajima, D.I. Santiago
and especially P.A. Lee for helpful discussions, and  F. Kr\"{u}ger for help with the
pictures. This work is supported by the Dutch Science Foundation NWO/FOM.


\begin{thebibliography}{99}

\bibitem{hilgenkamp} J.R. Kirtley \textit{et al.}, Nature Physics \textbf{2}, 190 (2006);
H.J.H. Smilde \textit{et al.}, Phys. Rev. Lett. \textbf{95}, 257001 (2005).

\bibitem{tajima} T. Masui \textit{ et al.}, Phys. Rev. B \textbf {68}, 060506 (2003);
R. Nemetschek \textit{ et al.}, Eur. Phys. J. B \textbf {5}, 495 (1998).

\bibitem{klemm} R.A. Klemm, Philos. Mag. \textbf{85}, 801, (2005);
R.A. Klemm, Philos. Mag. \textbf{86}, 2811, (2006).

\bibitem{wenlee} X.G. Wen and P.A. Lee, Phys. Rev. Lett. \textbf{ 76}, 503 (1996)

\bibitem{wenpsg} X.-G. Wen, Phys. Rev. B \textbf{65}, 165113 (2002).

\bibitem{hellberg} C. Stephen Hellberg and E. Manousakis, Phys. Rev. Lett. \textbf{78}, 4609
(1997).

\bibitem{kivelsonPS} V.J. Emery, S.A. Kivelson, H.G. Lin, Phys. Rev. Lett. \textbf{64}, 475
(1990).

\bibitem{seamus} T. Hanaguri \textit{ et al.}, Nature \textbf{ 430}, 1001 (2004).


\bibitem{zaanen} J. Zaanen and O. Gunnarson, Phys. Rev. B \textbf{ 40}, 7391 (1989).

\bibitem{white} Steven R. White and D.J. Scalapino, Phys. Rev. B \textbf{61}, 6320 (2000).

\bibitem{kivelsonstripe} E. Arrigoni \textit{et al.}, Phys. Rev. B \textbf{65}, 134503 (2002).


\bibitem{fatfollowuppapers}
P.A. Lee, N. Nagaosa and X.G. Wen, Rev. Mod. Phys. \textbf{ 78}, 17 (2006); P.A. Lee, N.
Nagaosa, T.-K. Ng and X.G. Wen, Phys. Rev. B \textbf{ 57}, 6003 (1998).


\bibitem{affleck} I. Affleck and J.B. Marston, Phys. Rev. B \textbf{37}, 3774 (1988);
T.C. Hsu, J.B. Marston and I. Affleck, Phys. Rev. B \textbf{43}, 2866 (1991).

\bibitem{kohn} W. Kohn and L.J. Sham, Phys. Rev. \textbf{140}, A1133 (1965).

\bibitem{simann} C.H. Chung, J.B. Marston and R.H. McKenzie, J. Phys.: Condens. Matter \textbf{
13}, 5159 (2001); W.H. Press \textit{ et al.}, \textit{ Numerical Recipes in C: The Art of
Scientific Computing}, 2nd. Ed., pp. 451-455, (Cambridge University press, New York, N.Y.).

\bibitem{kotliar} Z. Wang, G. Kotliar and X.-F. Wang, Phys. Rev. B \textbf{8690}, (1990).

\bibitem{tobepublished} B.W.A. Leurs and J. Zaanen, details to appear in a longer paper which
is in preparation.

\bibitem{fujimori} A. Ino \textit{ et al.}, Phys. Rev. Lett. \textbf{ 79}, 2101 (1997).

\bibitem{hybertsen} M.S. Hybertsen \textit{et al.}, Phys. Rev. B \textbf{41}, 11068 (1990).

\end{thebibliography}
\end{document}